# Local electric field induced spin photocurrent in ReS$_2$

*Yang Zhang[1,2], Yu Wang[4], Yu Liu[1,2], Xiao-Lin Zeng[1,2], Jing Wu[1,2], Jin ling Yu[3], Tian-Jun Cao[4], Shi-Jun Liang[4], Feng Miao[4\*] and Yong-Hai Chen[1,2\*]*

[1]Key Laboratory of Semiconductor Materials Science, Institute of Semiconductors, Chinese Academy of Sciences, Beijing Key Laboratory of Low Dimensional Semiconductor Materials and Devices, Beijing 100083, China

[2]Center of Materials Science and Optoelectronics Engineering，University of Chinese Academy of Sciences, Beijing 100049, China；

[3]Institute of Micro/Nano Devices and Solar Cells, School of Physics and Information Engineering, Fuzhou University, Fuzhou, China

[4]National Laboratory of Solid State Microstructures, School of Physics, Collaborative Innovation Center of Advanced Microstructures, Nanjing University, Nanjing 210093, China

\*Corresponding author: yhchen@semi.ac.cn; miao@nju.edu.cn.

## Abstract

Circular polarized photocurrent is observed near the electrodes on a few-layer ReS2 sample at room temperature. For both electrodes, the spatial distribution of the circular polarized photocurrent shows a feature of two wings, with one positive and the other negative. We suggest that this phenomenon arises from the inverse spin Hall effect due to local electric field near the electrode. Bias voltage that modulates this field further controls the sign and magnitude of the inverse spin Hall effect photocurrent. Our research shows that electric field near electrodes has a significant impact on spin transmission operation, hence it could be taken into account for manufacturing spintronic devices in future.

# Introduction

Over the past decade, various electrical, magnetic and optoelectronic methods to decouple the spin degree of freedom from charge degree of freedom have been demonstrated[1, 2, 3, 4]. Manipulating the spin degree of freedom, however, is associated with long-standing issues[5, 6, 7]. The use of circular polarized light for spin injection and regulation has been widely studied in many materials. The resulting effects of circular photogalvanic effect (CPGE), anomalous circular photogalvanic effect (ACPGE) and photo-induced anomalous Hall effect (PAHE) etc. are widely found in various semiconductors and topological materials[8, 9, 10, 11, 12, 13, 14, 15, 16, 17]. CPGE is traditionally gyrotropic optical media with a strong spin-orbit coupling, so that the effect is ordinarily controlled by angular momentum selection rules for excitation with circular polarized light. ACPGE and PAHE are derived from the inverse spin Hall effect of spin-polarized photo-generated carrier diffusion and drift under electric field respectively[13, 16, 17]. All of the above effects are the study of the circular polarized photocurrent of the materials. However, as the device is miniaturized, the boundary scattering, geometric shape, and contact with the electrodes can affect the transport of photocurrent[18, 19]. Therefore, focusing the light spot on the micrometer scale, using the microscopic imaging technique to obtain the photocurrent mapping of the sample becomes an effective means to study the local photocurrent of the device.

In this paper, we use microscopic imaging technology to obtain the ordinary photocurrent mapping and circular polarized photocurrent mapping of the few-layer $ReS_2$ sample. $ReS_2$ has attracted a lot attention due to its unique properties, such as anisotropy, ultra-high photo responsivity[20, 21]. Unlike most transition metal dichalcogenide monolayers (TMDs) with stable hexagonal phases, $ReS_2$ forms a distorted 1T structure with triclinic symmetry[22, 23, 24]. Also, spin-orbit coupling consists in $ReS_2$[25, 26]. By comparing the circular polarized photocurrent with the ordinary photocurrent, we find that the spatial distribution of the circular polarized photocurrent shows a feature of two wings, with one positive and the other negative.

Further experiment and analysis demonstrate that this strange phenomenon is derived from the inverse spin Hall effect of the local electric field near the electrodes. Moreover, this local inverse spin Hall effect can also be controlled by an external electric field. The discovery provides a new direction for the development of spintronics devices.

# Result

**Sample information and spectral properties**

Figure 1a shows the sample structure and experiment schematic. In our experiments, the sample is a Hall bar structure with two pairs of electrodes, in which the small electrodes are measuring electrodes and the strip electrodes used to apply voltage. A 532 nm laser with a spot size of around 1μm is shining to the surface of the sample perpendicularly and the sample is moved in by a two-dimensional nano-translation stage with a step size of 1μm. The incident light goes through a polarizer and a photo-elastic modulator (PEM), of which the peak retardation is set to λ/4, to yield a modulated circular polarized light with a fixed modulating frequency at 50 KHz (1F). An optical chopper with a chopping frequency of 223 Hz is used. The circular polarized photocurrent and ordinary photocurrent are collected through the two small electrodes by two lock-in amplifiers with the synchronization frequencies set to 50 KHz and 223 Hz, respectively. The detailed experimental configuration is in supplementary. In the photoluminescence (PL) and photocurrent measurements, a 532 nm laser is used as the excitation source, which generates inter-band transitions of electrons. The PL and photocurrent spectra are shown in Fig. 1c.

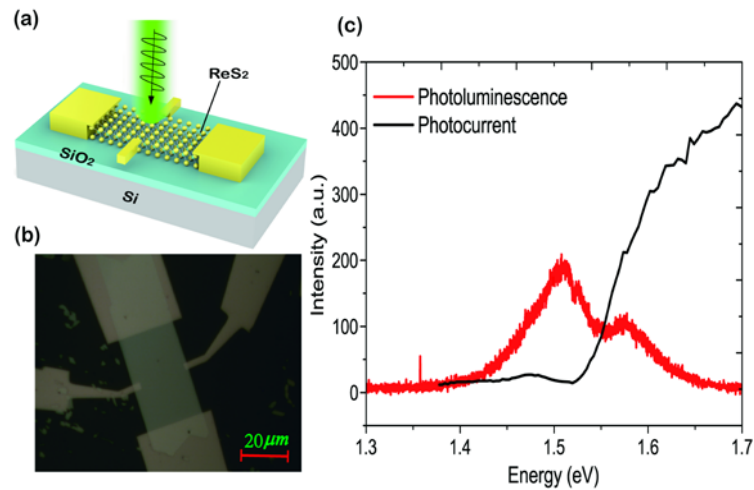

**Figure 1 | Sample information and spectral.** (**a**) Sample structure and experiment schematic. The long strip ReS$_2$ is placed on the SiO$_2$ substrate. Four metal electrodes are evaporated on the sample to form a hall bar structure. The pair of small electrodes are measuring electrodes and the large strip electrodes are used to apply transverse voltage. A 532 nm laser is illuminated on the surface of the sample. (**b**) The microscopic image of the sample. (**c**) The photoluminescence and photocurrent spectra of the sample. The red and black curve is the photoluminescence and photocurrent spectra, respectively. The 532 nm laser enables inter-band excitation of the sample in the PL and photocurrent spectra.

**Ordinary photocurrent and circular polarized photocurrent**

The scanning range of the two-dimensional translation stage is 50μm×50μm. Figure 2a and 2c are ordinary and circular polarized photocurrent mapping of the sample respectively. It can be seen that the photocurrent signal becomes larger with the light spot is closer to the electrode (as shown in Fig. 2a). The photocurrent signal of the sample is reversed due to the photovoltaic effect around the measuring electrodes. Fig. 2b and 2d show the relationship of ordinary and circular polarized photocurrent with position in x direction when the light spot scan in the line of y=20μm. By comparing the ordinary photocurrent and the circular polarized photocurrent, we are surprised to find that the circular polarized photocurrent signal is opposite on different side of the same measuring electrode and circular polarized photocurrent signal is same on the same side of the different measuring electrodes. The area that shows circular polarized photocurrent is much larger than the size of the electrode, which may indicate that the circular polarized photocurrent signal comes from the sample rather than the metal electrode.

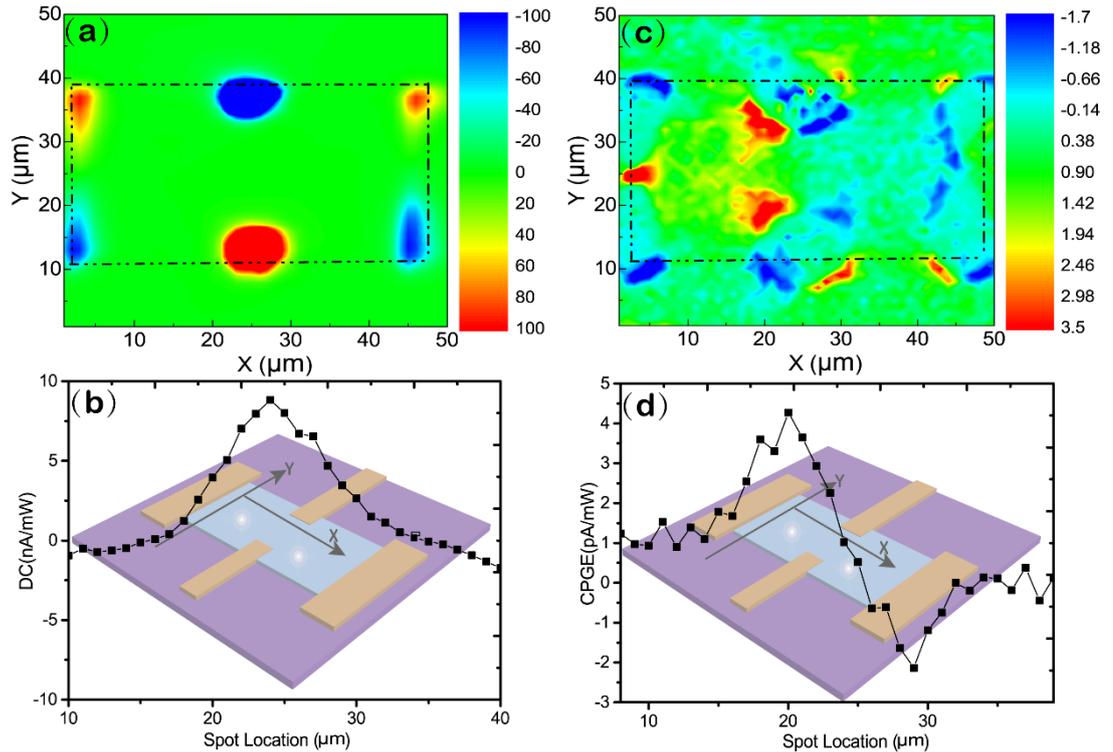

**Figure 2 | Ordinary photocurrent and circular polarized photocurrent.** (a) The ordinary photocurrent mapping of the sample. The black dotted part is the sample area and the scope of the sweep is 50μm×50μm. (b) The relationship of ordinary photocurrent with position in x direction when the light spot is in the y=20μm position. (c) The circular polarized photocurrent mapping of the sample. (d) The relationship of circular polarized photocurrent with position in x direction when the light spot is in the y=20μm position.

**Ordinary photoconductivity current**

In our study, the photoconductivity current is defined as the photocurrent signal measured under an external applied bias voltage. Fig. 3a and 3c show the ordinary photoconductive mapping of the sample when a bias of -0.3 and 0.3 V are applied on the measuring electrodes (small electrodes), respectively. Compared with Fig. 3b, we can obviously find that the photocurrent signal near the *+y (or -y)* electrode increases when a negative (or positive) bias voltage is applied, and when a negative bias voltage is applied, the area of the photocurrent signal in *+y* direction (red) will decrease and that in *–y* direction (blue) will increase. Fig. 3d, 3e and 3f show the relationship of ordinary photoconductivity with position in *y* direction when the light spot scans in the line of *x*=25 μm and a bias of -0.3V, 0V, 0.3V are applied on the measuring electrodes, respectively. The peak photocurrent changes approximately linearly with the applied bias voltage.

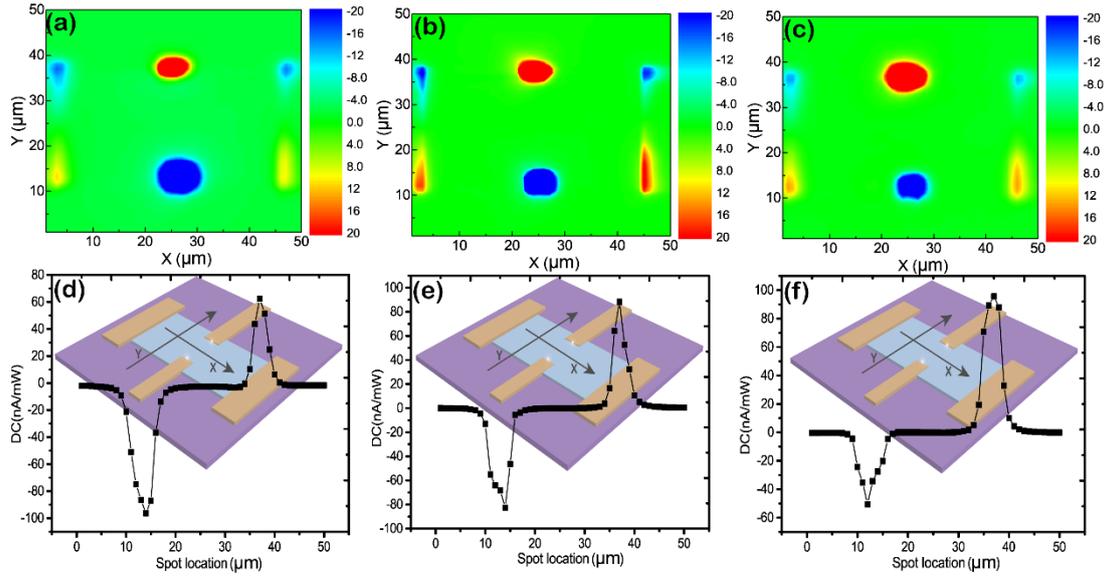

**Figure 3 | Ordinary photoconductivity current. (a)-(c) are** the ordinary photoconductivity current mapping of the sample when a bias voltage of -0.3, 0 and 0.3 V are applied on the measuring electrodes, respectively. **(d)-(f) show** the relationship of ordinary photoconductivity current with the position in *y* direction when a bias voltage of -0.3, 0 and 0.3 V are applied on the measuring electrodes, respectively, and the light spot is moving on the line of *x*=25 μm.

## Ordinary photocurrent and circular polarized photocurrent under the transverse electric field

The pair of large strip electrodes are used to apply transverse electric field. The ordinary photocurrent and circular polarized photocurrent under the transverse electric field are shown in Fig. 4. Figs. 4a and 4c are the ordinary photocurrent and circular polarized photocurrent mapping of the sample, respectively, when the transverse voltage is 0.3 V. Figs. 4b and 4d are the ordinary photocurrent and circular polarized photocurrent mapping of the sample, respectively, when the transverse voltage is -0.3V. By comparing the ordinary photocurrent (Fig. 4a and 4b) under different transverse electric field, we find that the photocurrent signal near the corresponding large strip electrode increases when a bias voltage in opposite direction is applied and the relationship is consistent with the ordinary photoconductivity current shown in Fig. 3.

However, it is interesting that change relationship of the circular polarized photocurrent under different transverse electric field. We have found that the circular polarized photocurrent signal is opposite on different side of the same measuring electrodes and circular polarized photocurrent signal is the same on the same side of the different measuring electrodes as shown in Fig.2. Now we find that the signal of the circular polarized photocurrent will change correspondingly with the transverse voltage. Coincidentally, the active area of the circular polarized photocurrent also change with the corresponding transverse voltage in different directions (as shown in Fig. 4c and 4d). We infer that this phenomenon comes from the inverse spin Hall effect due to the local electric field near the electrode and we will discuss it in detail in the next part. Figs. 4e and 4f show the schematic of the local electric field around the electrodes under transverse electric field in different directions.

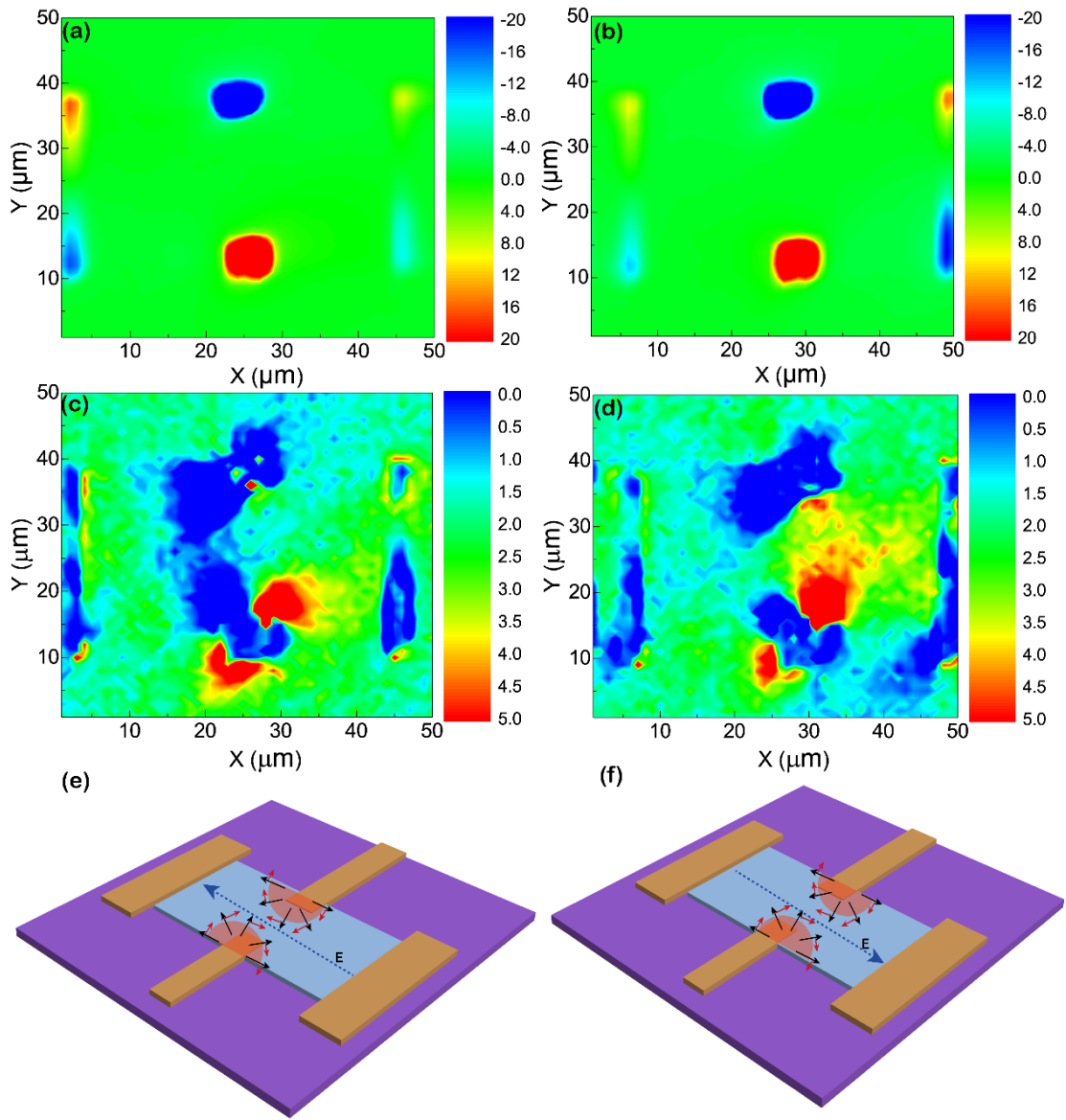

**Figure 4 | Ordinary photocurrent and circular polarized photocurrent under the transverse electric field. (a) and (b) are the** ordinary photocurrent mapping of the sample when the transverse voltage is -0.3 and 0.3 V, respectively. **(c) and (d) are** the circular polarized photocurrent mapping of the sample when the transverse voltage is -0.3 and 0.3 V, respectively. **(e) and (f) are** the schematic of local electric field around the electrode under a transverse voltage of 0.3 and -0.3V, respectively.

# Discussion

**Model of inverse spin Hall effect due to local electric field around the electrodes in ReS₂**

Contact barrier exists due to the different Fermi levels of the metal electrode and the sample. The spatial dependence $E(x)$ of the electric field strength is

$$E(x) = -\frac{eN_D}{\varepsilon_S}(\omega_0 - x) = E_m + \frac{eN_D}{\varepsilon_S}x$$

where $\varepsilon_S = \varepsilon_0 \varepsilon_r$ is the static dielectric constant of the semiconductor and $\omega_0$ being the depletion layer width at zero bias. What's more, the maximum field strength is $E_m = -\frac{eN_D}{\varepsilon_S}\omega_0$ at $x=0$, from the condition $\varphi(\omega_0) = 0$, we obtain $\omega_0$ as $\omega_0 = \sqrt{\frac{2\varepsilon_S}{eN_D}V_{bi}}$. Here we defined the $N_D$ is the charge density of the ReS₂ and $V_{bi}$ is the built-in potential. The detailed derivation process is in the supplementary materials (part II).

The black arrow near the electrodes in Fig. 5 represents the local electric field owing to contact barriers. Circular polarized light produces spin-polarized photo-generated carriers and the spin-polarized photo-generated carriers undergo directional drift due to the directional drive of the local electric field. As a result, there is a spin current flowing in the same direction as the local electric field near the electrode. Due to the inverse spin Hall effect, a corresponding charge current flowing in the vertical direction of the local electric field will be generated (the red arrow in Fig. 5). This inverse spin Hall photocurrent signal is opposite on different sides of the same measuring electrode. We also did some simulations with current sources in different directions and found that they are consistent with our experimental results. The detailed calculation process is shown in supplementary materials（as shown in Figure S2）.

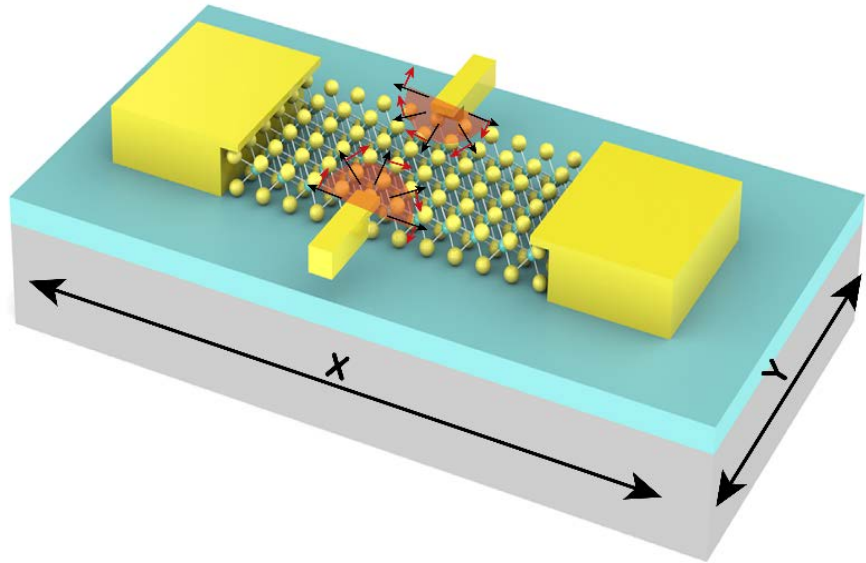

**Figure 5 | Schematic of the inverse spin Hall effect due to local electric field around the electrode.** The black arrow represents the local electric field and red arrow represents the circular polarized photocurrent due to inverse spin Hall effect. Translucent red area is the role of local electric field near the electrode.

**The relationship of circular polarized photocurrent due to inverse spin Hall effect around the electrodes in ReS$_2$ with the transverse electric field.**

From previous studies[14], we know that the circular polarized photocurrent due to inverse spin Hall effect shows linear dependence on the applied transverse voltage[14]. We also find that circular polarized photocurrent due to inverse spin Hall effect around the electrodes in ReS$_2$ has the same relationship. Fig. 6a shows the relationship of circular polarized photocurrent with position in $x$ direction when the light spot scan in the line of y=20 μm position and the transverse voltage is 0V, ±0.1V and ±0.3V, respectively. Fig. 6b and 6c show relationship of circular polarized photocurrent with the transverse voltage when the light spot fixed to a specific area on the left and right side of the electrode, respectively. The light spot positions are indicated by the white spot in Fig. 6b and 6c.This linear relationship with the transverse voltage can be explained as follows: The local electric field near the electrode can be changed by the transverse voltage，and consequently, the circular polarized photocurrent also changed with the transverse voltage. Specifically, when a positive (or negative) transverse voltage is applied, the transverse local electric field on one side of the electrode is strengthened (or weakened), and the local transverse electric field on the other side is correspondingly weakened (or strengthened). So the value of circular polarized photocurrent is 0.9 pA/mw on the left side of the electrode and 2.15 pA/mw on the right side of the electrode when transverse voltage is -0.3 V. Simultaneously, the active area of the circular polarized photocurrent also increases or decreases with the corresponding transverse voltage in different directions. The blue region and the red region become larger and smaller corresponding to the applied transverse electric field. (as shown in Fig. 4c and 4d).

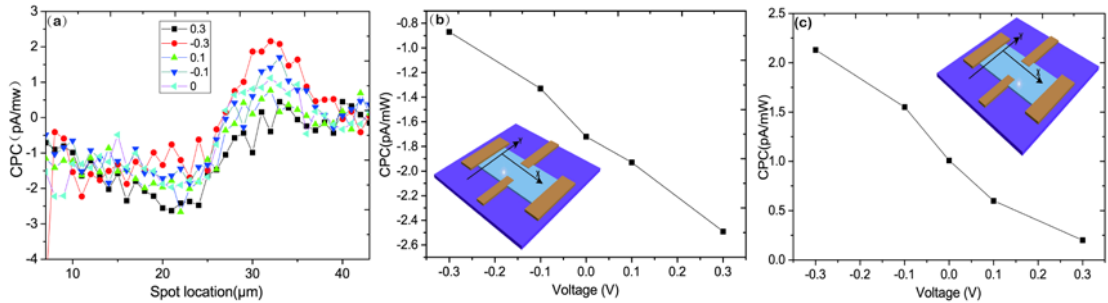

**Figure 6 | Circular polarized photocurrent with the transverse electric field.** (a) The relationship of circular polarized photocurrent with position in *x* direction when the light spot is in the y=20 μm position and the transverse voltage is 0V, ±0.1V and ±0.3V, respectively. (b) and (c) show the relationship of circular polarized photocurrent with the transverse voltage when the light spot fixed to a specific area on the left and right side of the electrode, respectively. The light spot positions are indicated by the white spot

In conclusion, we have systematically studied the ordinary photocurrent and circular polarized photocurrent of few-layer ReS$_2$, a semiconducting 2D TMD with a distorted 1T structure. By comparing the ordinary photocurrent and the circular polarized photocurrent, we are surprised to find that the circular polarized photocurrent signal is opposite on different sides of the same measuring electrode, which is caused by the inverse spin Hall effect of the local electric field. Besides, by applying a transverse electric field, we can tune the local electric field and the circular polarized photocurrent due to inverse spin Hall effect of the local electric field. This phenomenon is ubiquitous and we also find it in few-layer WTe$_2$ device (the detailed experimental results are shown in supplementary materials part IV). Because the effect can be engineered by a combination of shape and applied fields, many applications, as well as the ability to encode more information in a device by using the spin degrees of freedom, are possible.

# Method

**Experimental setup for photocurrent measurements**

After the laser beam from the source (wavelength 532 nm from a 300 mW solid state laser, 2% output power stability) is spatially filtered, it passes through a polarizer (set at 0) and photo-elastic modulator (PEM, set at 45°). Then the beam, 1μm in estimated diameter, is splitted by a polarization preserved 50/50 beam splitter down into a long working distance objective lens (Mitutoyo 100X, NA = 0.95), and is focused on the sample taken on a piezoelectric ceramic electric translation (PCET) stage which can move in two dimensions in plane. An optical chopper with the rotation frequency of 223 Hz is used. The circular polarized photocurrent and ordinary photocurrent are collected through the two circle electrodes by two lock-in amplifiers with the synchronization frequencies set to be 50 KHz and 223 Hz, respectively. The detail setup of our experiment is in the supplementary materials (part I).

**Materials and devices**.

Single crystals of $ReS_2$ were grown by the same Br2-assisted chemical vapour transport method. We used a standard mechanical exfoliation method to isolate few-layer $ReS_2$ films. A conventional electron-beam lithography process (FEI F50 with Raith pattern generation system) followed by standard electron-beam evaporation of metal electrodes (typically 5 nm Ti/50nm Au) was used to fabricate few-layer $ReS_2$ sample.

Reference

1. Zutic I, Fabian J, Das Sarma S. Spintronics: Fundamentals and applications. *Reviews of Modern Physics* **76**, 323-410 (2004).

2. Zeng HL, Dai JF, Yao W, Xiao D, Cui XD. Valley polarization in MoS2 monolayers by optical pumping. *Nature Nanotechnology* **7**, 490-493 (2012).

3. Xiao D, Liu GB, Feng WX, Xu XD, Yao W. Coupled Spin and Valley Physics in Monolayers of MoS2 and Other Group-VI Dichalcogenides. *Physical Review Letters* **108**, (2012).

4. Kato YK, Myers RC, Gossard AC, Awschalom DD. Observation of the spin hall effect in

semiconductors. *Science* **306**, 1910-1913 (2004).

5. Bhardwaj RG, Lou PC, Kumar S. Spin Seebeck effect and thermal spin galvanic effect in Ni80Fe20/p-Si bilayers. *Applied Physics Letters* **112**, (2018).

6. Ganichev SD*, et al.* Spin-galvanic effect. *Nature* **417**, 153-156 (2002).

7. Wunderlich J*, et al.* Spin Hall Effect Transistor. *Science* **330**, 1801-1804 (2010).

8. Okada KN*, et al.* Enhanced photogalvanic current in topological insulators via Fermi energy tuning. *Physical Review B* **93**, (2016).

9. Zhang S*, et al.* Generation of Rashba Spin-Orbit Coupling in CdSe Nanowire by Ionic Liquid Gate. *Nano Letters* **15**, 1152-1157 (2015).

10. Yu JL*, et al.* Temperature dependence of spin photocurrent spectra induced by Rashba- and Dresselhaus-type circular photogalvanic effect at inter-band excitation in InGaAs/AlGaAs quantum wells. *Optics Express* **23**, 27250-27259 (2015).

11. Yu JL, Cheng SY, Lai YF, Zheng Q, Chen YH. Spin photocurrent spectra induced by Rashba- and Dresselhaus-type circular photogalvanic effect at inter-band excitation in InGaAs/GaAs/AlGaAs step quantum wells. *Nanoscale Research Letters* **9**, (2014).

12. Olbrich P*, et al.* Room-temperature high-frequency transport of dirac fermions in epitaxially grown Sb2Te3- and Bi2Te3-based topological insulators. *Phys Rev Lett* **113**, 096601 (2014).

13. Zhu LP*, et al.* Excitation wavelength dependence of the anomalous circular photogalvanic effect in undoped InGaAs/AlGaAs quantum wells. *Journal of Applied Physics* **115**, (2014).

14. Mei F*, et al.* Spin transport study in a Rashba spin-orbit coupling system. *Scientific Reports* **4**, (2014).

15. Peng XY*, et al.* Anomalous linear photogalvanic effect observed in a GaN-based two-dimensional electron gas. *Physical Review B* **84**, (2011).

16. Tang CG, Chen YH, Liu Y, Wang ZG. Anomalous-circular photogalvanic effect in a GaAs/AlGaAs two-dimensional electron gas. *Journal of Physics-Condensed Matter* **21**, (2009).

17. He XW*, et al.* Anomalous photogalvanic effect of circularly polarized light incident on the two-dimensional electron gas in Al(x)Ga(1-x)N/GaN heterostructures at room temperature. *Physical Review Letters* **101**, (2008).

18. Xie L, Cui XD. Manipulating spin-polarized photocurrents in 2D transition metal dichalcogenides. *Proceedings of the National Academy of Sciences of the United States of*

*America* **113**, 3746-3750 (2016).

19. Chen X, Yan TF, Zhu BR, Yang SY, Cui XD. Optical Control of Spin Polarization in Monolayer Transition Metal Dichalcogenides. *Acs Nano* **11**, 1581-1587 (2017).

20. Liu EF, *et al.* Integrated digital inverters based on two-dimensional anisotropic ReS2 field-effect transistors. *Nature Communications* **6**, (2015).

21. Jing QH, *et al.* Ultrasonic exfoliated ReS2 nanosheets: fabrication and use as co-catalyst for enhancing photocatalytic efficiency of TiO2 nanoparticles under sunlight. *Nanotechnology* **30**, (2019).

22. Zereshki P, Yao P, He D, Wang Y, Zhao H. Interlayer charge transfer in ReS2/WS2 van der Waals heterostructures. *Physical Review B* **99**, (2019).

23. Oliva R, *et al.* Pressure dependence of direct optical transitions in ReS2 and ReSe2. *Npj 2d Materials and Applications* **3**, (2019).

24. Luo M, *et al.* Gate-tunable ReS2/MoTe2 heterojunction with high-performance photodetection. *Optical and Quantum Electronics* **51**, (2019).

25. Webb JL, Hart LS, Wolverson D, Chen CY, Avila J, Asensio MC. Electronic band structure of ReS2 by high-resolution angle-resolved photoemission spectroscopy. *Physical Review B* **96**, (2017).

26. Kim BS, Kyung WS, Denlinger JD, Kim C, Park SR. Strong One-Dimensional Characteristics of Hole-Carriers in ReS2 and ReSe2. *Scientific Reports* **9**, (2019).

**Acknowledgement:**

The work is supported by the National Basic Research Program of China (Grants No. 2015CB921503), the National Natural Science Foundation of China (Grants No. 61474114, No. 11574302, 61627822, 11704032), National Key Research and Development Program (Grant No. 2018YFA0209103 No. 2016YFB0402303 and No. 2016YFB0400101).